\documentclass[8.5pt,twoside,twocolumn]{article}
\usepackage[super,sort&compress,comma]{natbib} 
\usepackage{mhchem}
\usepackage{times,mathptm}
\usepackage{balance} 
\usepackage{color}

\usepackage{epstopdf}  % JA: to get it to work for m on MAC...

\usepackage{graphicx} %eps figures can be used instead
\usepackage{lastpage}
\usepackage[format=plain,justification=raggedright,singlelinecheck=false,font=small,labelfont=bf,labelsep=space]{caption} 
\usepackage{fancyhdr}
\usepackage{epic}
\newcommand{\um}{\:\mu\text{m}}

\newcommand{\BS}{{\it B.\,subtilis}}

\begin{document}

\thispagestyle{plain}
\fancypagestyle{plain}{
%\fancyhead[L]{\includegraphics[height=8pt]{headers/LH.pdf}}
%\fancyhead[C]{\hspace{-1cm}\includegraphics[height=20pt]{headers/CH.pdf}}
%\fancyhead[R]{\includegraphics[height=10pt]{headers/RH.pdf}}
\renewcommand{\headrulewidth}{1pt}}
\renewcommand{\thefootnote}{\fnsymbol{footnote}}
\renewcommand\footnoterule{\vspace*{1pt}% 
\hrule width 3.4in height 0.4pt \vspace*{5pt}}

\makeatletter 
\renewcommand\@biblabel[1]{#1}            
\renewcommand\@makefntext[1]% 
{\noindent\makebox[0pt][r]{\@thefnmark\,}#1}
\makeatother 
\renewcommand{\figurename}{\small{Fig.}~}
%\sectionfont{\large}
%\subsectionfont{\normalsize} 

\fancyfoot{}
%\fancyfoot[LO,RE]{\vspace{-6pt}\includegraphics[height=8.5pt]{LF.pdf}}
%\fancyfoot[CO]{\vspace{-6.5pt}\hspace{11.4cm}\includegraphics{RF.pdf}}
%\fancyfoot[CE]{\vspace{-6.6pt}\hspace{-12.7cm}\includegraphics{RF.pdf}}
%\fancyfoot[RO]{\footnotesize{\sffamily{1--\pageref{LastPage} ~\textbar  \hspace{2pt}\thepage}}}
%\fancyfoot[LE]{\footnotesize{\sffamily{\thepage~\textbar\hspace{4.4cm} 1--\pageref{LastPage}}}}
\fancyhead{}
\renewcommand{\headrulewidth}{1pt} 
\renewcommand{\footrulewidth}{1pt}
\setlength{\arrayrulewidth}{1pt}
\setlength{\columnsep}{6.5mm}
\setlength\bibsep{1pt}

\twocolumn[
  \begin{@twocolumnfalse}
\noindent\LARGE{\textbf{Colloids in a bacterial bath: simulations and experiments}}
\vspace{0.6cm}

\noindent\large{\textbf{Chantal Valeriani,$^{\ast}$\textit{$^{a}$} Martin Li\textit{$^{a}$}, John Novosel\textit{$^{a}$},
Jochen Arlt\textit{$^{a}$} and Davide Marenduzzo$^{a}$ }}\vspace{0.5cm}
%%Please note that \ast indicates the corresponding author(s) but no footnote text is required. 

%\noindent\textit{\small{\textbf{Received Xth XXXXXXXXXX 20XX, Accepted Xth XXXXXXXXX 20XX\newline
%First published on the web Xth XXXXXXXXXX 200X}}}

\noindent \textbf{\small{DOI: 10.1039/b000000x}}
\vspace{0.6cm}
%Please do not change this text.

\noindent \normalsize{    
We present a joint experimental and computational study of the effect of bacterial motion on micron-scale colloids contained in a two-dimensional suspension of \BS. With respect to previous work using {\it E. coli}, here we introduce a novel experimental set-up that allows us to realise a two-dimensional bacterial suspension insensitive to either evaporation or fluid flow.
By analysing the mean square displacements of both bacteria and colloids, we confirm the existence of a crossover from super-diffusive behaviour at short time scales to normal diffusion at longer times.
We also study the same two-dimensional system by means of numerical simulations, using a suspension of self-propelled dumbbells or the Vicsek model, which has been previously used to study the dynamics of active particles.
Our numerical results obtained with both models are in broad agreement with the experimental trends, but only the dumbbell simulations can match the experimental data quantitatively. The level of agreement we find suggest that steric interactions due to collisions are important players in determining collective motion of the bacterial bath, and should complement hydrodynamic interactions in experiments. }
\vspace{0.5cm}
 \end{@twocolumnfalse}
  ]

\section{Introduction}\label{sec:introduction}

%Footnotes
\footnotetext{ $^\textit{a}$ \it{SUPA School of Physics and Astronomy, University of Edinburgh, Mayfield Road, Edinburgh EH9 3JZ, United Kingdom.}}

\footnotetext{\dag~Electronic Supplementary Information (ESI) available: Video of a $9\%$ suspension of {\it B. Subtilis} seeded with a 
small percentage of fluorescently labelled cells. Epi-fluorescence and standard bright-field illumination are combined to visualise 
all the suspended cells, but only the fluorescent cells are used for the particle tracking. The field of view is about $225\, \mu m$ 
by $168\,\mu m$ in dimension and the video was recorded at a rate of 10 frames per second. See DOI: 10.1039/b000000x/}

%Please use \dag to cite the ESI in the main text of the article.
%If you article does not have ESI please remove the the \dag symbol from the title and the above footnotetext.
%additional addresses can be cited as above using the lower-case letters, c, d, e... If all authors are from the same address, no letter is required
%\footnotetext{\ast~Additional footnotes to the title and authors can be included \emph{e.g.}\ `Present address:' or `These authors contributed equally 
%to this work' as above using the symbols: \ddag, \textsection, and \P. Please place the appropriate symbol next to the author's name and include a \texttt{\textbackslash footnotetext} entry in the the correct place in the list.}

Schools of fish, microbial colonies and bacterial suspensions provide intriguing examples of self-driven soft matter which is far from equilibrium 
even in steady state~\cite{sriram,menon}. Their dynamics display spectacular pattern formation and collective behaviour and are dominated by long-range 
correlations, as can be observed in flocks of birds or schools of fish~\cite{ortiz}. Likewise, bacteria can swarm and concentrated films of swimming bacteria 
facilitate complicated self-driven flows which are several times larger in magnitude than swimming velocities of individuals~\cite{goldstein1}: this
 phenomenon is sometimes referred to as ``bacterial turbulence''. Understanding the mechanism underlying ordering and patterning in suspensions of self-driven systems has now become a
topical question in soft matter and biological physics. 

Besides providing an intriguing playground for ideas in non-equilibrium physics, self-propelled living matter may well provide templates
for the next generation of artificial smart materials, which can be made, for instance,
to swim in a controlled way in aqueous solutions. `First-principle' physics
studies of bacterial suspensions and colonies have already been successful at
providing possible mechanisms for the patterning, for instance in chemotactic strains~\cite{mike}.

Dombrowski and coworkers have experimentally studied cell-driven motions in droplets of {\it Bacillus~subtilis}~\cite{goldstein1}. 
Pedley and Kessler observed the formation of convection patterns in sessile drops, due to Rayleigh-Taylor instabilities induced by the motion of cells towards the free surface, richer in oxygen concentration~\cite{kessler1}. In pendant drops, the authors observed flow patterns which originated from hydrodynamics interactions. %JA

In a seminal work, Wu and Libchaber~\cite{Wu:2000/a} measured the diffusion of colloidal tracers in a bacterial bath, 
and found that these behaved superdiffusively at early times, and exhibited a crossover to a diffusional regime at later times. 
The crossover between the two different regimes occurred at about $1\, $s, which is about 10 times smaller than the Brownian time over 
which a single colloidal sphere would diffuse a distance equal to its size. 
%The likely explanation for this phenomenology was that at early times colloids were pushed around by swimmers, and this led to their superdiffusive behaviour. 
%Then, due to bacterial tumbling, many body hydrodynamics interactions or noise, both colloids' and bacteria's trajectories became essentially random, giving rise to diffusive dynamics. 
%JA:
The proposed explanation for this phenomenology is that at early times colloids are pushed around by swimmers which leads to their superdiffusive behaviour. 
Then, due to bacterial tumbling, many body hydrodynamics interactions or noise, both colloids' and bacteria's trajectories become essentially random, giving rise to diffusive dynamics. 
The effective diffusion coefficient associated with the latter regime is strikingly different in magnitude from the one 
which could be inferred from the Stokes-Einstein law for Brownian particles of the same size as bacteria. Still, the diffusion 
coefficient turned out to exhibit the same functional dependence on the colloid radius as in the Stokes-Einstein's expression~\cite{Wu:2000/a}.

In order to interpret their experimental data, Wu and Libchaber proposed a Langevin-type model where the ``collisional force'' 
acting on the colloids  due to the bacterial motion was exponentially correlated with a memory effect. In a subsequent work, 
Gregoire and coworkers \cite{comment:2001/a} pointed out that Langevin equations would predict a non-realistic ballistic behaviour at short time scales and suggested to use 
a model based on self-propelled particles, the Vicsek model, to study the collective dynamics of a swarm of bacteria\cite{Vic2:1997/a,chate3:1998/a,chate4:2001/a}.
In their Vicsek model, each self-propelled particle moves in the direction of its orientation, 
which is aligned with the average orientation of all particles in its neighbourhood, with some added noise. Particles also 
interact via a repulsive potential with passive beads suspended in the system. Using this model, Gregoire and coworkers 
qualitatively recovered the super-diffusive behaviour which was observed experimentally at short time scales. % JA, but not observed in Wu and Libchaber's numerical model.

However, it has now been realised that the existence of the crossover behaviour found by Wu and Libchaber is a generic phenomenon, to be expected on very general grounds in models of dynamics of a suspension of self-propelled particles, whether living or synthetic. 
%Indeed, more recently, Llopis and Pagonabarraga~\cite{isaac} performed lattice Boltzmann  simulations of polar self-propelled particles, 
%However, most of the latter only considered suspensions of self-propelled particles without dealing with a mixture of self-propelled particles and passive colloids, as in the original experiments in~\cite{Wu:2000/a}.
For instance, more recently a number of researchers have proposed numerical models for a suspension of self-propelled particles 
at low Reynolds numbers. Llopis and Pagonabarraga~\cite{isaac} performed lattice Boltzmann simulations of self-propelled particles,
 finding that ``hydrodynamic noise'', coming from the interaction between different swimmers, is already enough to lead to diffusive 
behaviour at large times, even in models not incorporating bacterial tumbling, and showing a crossover from ballistic to diffusive motion.
The latter result was confirmed by subsequent works, e.g. the Stokesian dynamics simulations presented in Ref.~\cite{mehandia}, and the numerical work by Ortiz and coworkers ~\cite{ortiz}. 
The latter work proposed another minimal model for a microscopic swimmer, capable of 
capturing the leading order (dipolar) far-field hydrodynamics interactions between swimmers. 
Each swimmer is modelled as a rigid dumbbell made of two beads connected by a rigid body; the drag acts on both beads and the flagellum exerts a constant force on one of 
the beads in the direction of the dumbbells' unit vector and an equal and opposite force on the fluid. Within this model, a crossover 
between super-diffusive behaviour at short time-scales and diffusive at longer time-scales was found. At high concentration the authors 
observed large-scale coherent fluid motion. However in none of these works a suspension containing both passive and active particles was considered  
as in the original experiments in Ref.~\cite{Wu:2000/a}.

We note, in passing, that another interesting theoretical framework within which to study suspensions of self-propelled particles is via a continuum theory, which has been successfully used to study microbial and bacterial fluids, for instance, in Ref.~\cite{tonertu1,tonertu2,ramaswamy1,ramaswamy2,baskaran}. It would be interesting to explore what predictions these theories yield as regards the existence and behaviour of the crossover in the dynamics of colloidal tracers in active fluids -- however here we do not address these questions as these models work in the limit of high density whereas our experiments are with relatively dilute suspensions. 

Notwithstanding the significant amount of both experimental and theoretical existing works, it still remains an 
open issue whether the early regime of the colloidal motion is ballistic or superdiffusive in nature. By considering a 
collection of non-interacting wild-type (run \& tumble)  bacteria, one might argue that the early time regime should
 correspond to simple runs and be ballistic and the later times, after one or more tumbles, should be diffusive, as the 
bacteria are effectively performing a random walk at large length and time scales. 
 However, most experiments show a 
superdiffusive behaviour at short time scales, with the mean square displacement behaving as a power law with an exponent $\alpha$
$1.5 < \alpha < 2$ \cite{Wu:2000/a}.
The reasons behind this peculiar behaviour are still not understood.
%Therefore, an interesting question one might ask is what is the reason for this peculiar behaviour. 
Furthermore, and importantly, we are not aware of a study which directly compares in detail experimental and simulation data.
Such a comparison would obviously be useful, and for instance 
allow to more stringently and quantitatively test the dynamical predictions coming from models of 
self-propelled particles. E.g., are propulsion and collisions enough to quantitatively match the data? Or is there
a large effect of hydrodynamic interactions coming from the (dipolar) forces which swimmers need to exert on the fluid 
in order to move? 

Our aim in this work is to attempt to answer these questions by combining results from experiments and simulations.
Experiments are performed on  a quasi two-dimensional bacterial suspension of {\it Bacillus~subtilis}
and our numerical simulations adopt two independent ways to represent self-propelled particles. 
On the one hand, we simulate the bacterial bath as Brownian Dynamics of self-propelled dumbbells, interacting via hard repulsion 
with passive colloids. This allows us to appropriately consider the effect of steric interactions, as well as of the rod-like 
shape of bacteria. 
On the other hand, we use the two-dimensional vectorial Vicsek model~\cite{chate4:2001/a,comment:2001/a} with the additional assumption that both self-propelled and colloidal particles have a finite size, represented by 
a short range hard-core repulsion, and  interact via a soft repulsion with the passive colloids embedded in the suspension. 

Our emphasis is on the comparison between simulations and experimental data in rather dilute bacterial films:  
in this way we can assess the accuracy of the various models proposed in literature for self-propelled particles 
and bacterial suspensions, with respect to (i) the existence and location of the crossover between superdiffusive and diffusive behaviour and 
(ii) the time series of the mean-square displacements of {\em both} the colloidal and 
the active particles. 
In our simulations we disregard fluid-mediated interactions between two dumbbells or self-propelled particles 
(although the orienting term coming from the Vicsek model may be thought as arising from hydrodynamic interactions among other things), 
and we will see that this is not crucial to either the existence of the crossover or its quantitative estimation. 
The mean-square displacements recorded from the simulations are only in semi-quantitative agreement with experiments, however the discrepancy 
may be due to other details which are not included in the simulations, such as the modelling of tumbling events in \BS\ 
(see Section~\ref{sec:experiments}, and the discussion of our results).

\section{Experimental set-up} \label{sec:experiments}
%%%\subsection{Cell preparation}

Similar to the pioneering work of Wu and Libchaber\cite{Wu:2000/a}  we experimentally study a quasi two-dimensional bacterial suspension. 
In our experiments we sandwich a dilute suspension of motile {\it Bacillus~subtilis}  in between two oxygen-plasma-treated 
glass cover-slides, creating thin samples (about $5-7 \um$ in  height) that can be imaged with high optical quality.
% that allows to visualise individual bacteria.
Particle tracking of individual particles is achieved by using a small fraction of fluorescently labelled tracers, either passive polystyrene beads or fluorescently labelled motile  bacteria.

\BS\ is a peritrichously flagellated bacterium often found in soil, with a general morphology characterised by a rod-shaped body approximately $4\um$ in length and $1\um$ in diameter (just after
cell division).
Although its swimming motility is not well characterised, it is assumed to perform a  ``run-and-tumble'' motion similar to that of the extensively studied {\it E.~coli}~\cite{BergEcoli}:
periods of swimming in straight trajectories (``runs'')  are separated by shorter ``tumbling'' events where cells randomly re-orientate their body.
Runs typically last for $\sim1\,$s during which  \BS\ swims with a speed of about $30 \um/\text{s}$. 
Although tumble events only last for few milliseconds, they randomise the direction of the trajectories.
As the combination of a large number of run-and-tumble events can be approximated by a random walk-like motion, it is natural to expect a diffusive-like motion over long time scales.

%Our experiments are conducted on diluted suspensions of {\it Bacillus~subtilis}  (DS1919) cells  (a peritrichously flagellated rod-shaped bacterium about $4 \times 1 \um$ in size) containing few colloidal tracers.
\subsection{Bacterial cultures and cell staining}

\BS\  (DS1919) cells are prepared from frozen stocks, where an overnight culture is first prepared for 16 hours in Luria-Bertani (LB) broth at $30^\circ$ Celsius 
%containing 10~grams of tryptone, 5~grams of yeast extract and 5~grams/l of NaCl, 
and then displaced in an orbital incubator (with a shaker speed of 200 rpm).
Next, \BS\ cells are incubated with 100 $\mu$g/ml of spectinomycin (Sigma-Aldrich, S4014) in order to allow for fluorescent labelling. 
We then harvest a $50 \mu$l sample from the overnight stock to inoculate 5~ml LB broth, which is incubated  for 5-6 hours in order to reach the mid-exponential growth phase of the bacteria. 
We then wash the sample once using a motility buffer containing 0.01 M of KPO$_4$, 0.067 M of NaCl and 
$10^{-4}$ M of EDTA (at pH 7.0) before re-suspending the bacteria  in motility buffer at the desired dilution. Notice that after the washing step bacterial growth stops.

\begin{figure}[h!]
\begin{center}
\includegraphics[width=3cm]{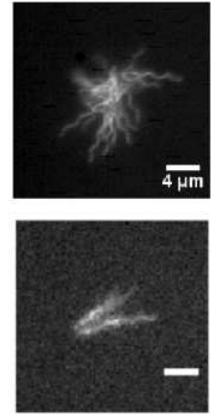}
\caption{\label{fig:staining} Fluorescently labelled \BS\ (DS1919),  where the dye  (Alexa Fluor 488 C5) only  stains the flagella. 
Top: Immobilised cell stuck to the cover-slip, highlighting the number of flagella and their shape. Bottom: a cell performing a run showing the formation of multiple flagella bundles. 
This peculiar flagellar arrangement during the run will be discussed in detail elsewhere~\cite{martin}.}
\end{center}
\end{figure}

To highlight individual swimming cells we adapt a flagella labelling  technique developed by Blair {\it et al.}~\cite{blair}.
The \BS\ cells are genetically modified to provide a high density of specific binding sites for a fluorescent dye molecules  along the length of its flagella filaments (for details of the genetic modification and staining procedure see Ref.~\cite{blair}).
The main advantage of this specific binding is that it allows to stain only the flagella filaments, thus avoiding light scattering from the cell body, unlike the one observed in {\it E.coli}, originated in standard staining techniques by non-specific binding~\cite{turner}.
%The genetic mutation consists of modifying one (non-essential)  exposed amino-acid of the outer sub-domain of the flagellin protein surface.  
%The side chain (residue) of the mutated amino-acid (amyE::Phag-hagT209C see reference \cite{blair}) is going to be the target for the specific binding of a fluorescent dye molecule (Alexa Fluor 488 C5 maleimide, Molecular Probes).
%Due to the high density of exposed residues in the flagellin peptide chain, the binding of the dye molecule is enough to visualise the flagellum in its entire length.
Figure~\ref{fig:staining} shows the results obtained after staining the flagella.

The staining technique does not only allow to extract the average number of flagella on the \BS\ cells used for our experiments, 
but also to estimate their length and characterise their general morphology. 
On average, our \BS\ cells have ($14.5 \pm 3.5$) flagella filaments per cell with an average filament length of $(8.5\pm 0.2)\um$. 
Bright field imaging shows that the cell body is $(6\pm 2)\um$ long (consistent with a population of cells at
different stages of growth) and an average run 
velocity is $(30\pm 10)\um/$s (around $40 \um/$s  in a very diluted sample).
It should be noted that labelling of the flagella provides much more information than required for the analysis presented in the following, 
for which much simpler labelling of the cell body would have been sufficient.
Fluorescent imaging of flagella of swimming cells reveals novel insight about the swimming characteristics of both individual \BS\ cells 
as well as their interaction in dense suspensions which will be presented elsewhere\cite{martin}.

For the experiments on thin films described below, samples consisted of a dilute suspension of (unlabelled) motile \BS\ cells seeded with a small fraction ($<5\%$) of fluorescent tracers:
 either fluorescently stained \BS\ cells as tracers for active swimmers or yellow-green fluorescent polystyrene beads of $4\um$ diameter (Invitrogen, F8859)  acting as passive tracers.

\subsection{Sample preparation and imaging}

\begin{figure}[h!]
\begin{center}
\includegraphics[width=9cm]{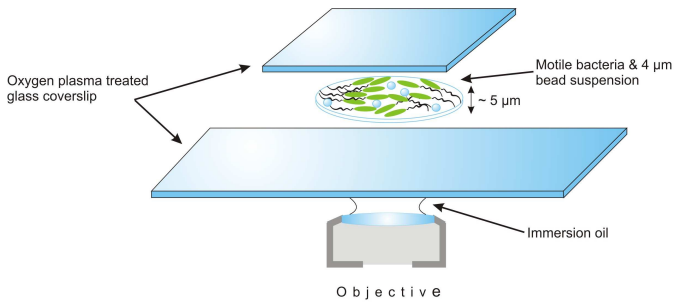}
\caption{\label{fig:setup} Schematic diagram of the experimental configuration, where a suspension of bacteria is confined in a layer about $5 \mu m$ thick between two oxygen-plasma treated glass cover-slips ($22\,$mm $ \times \; 50\,$mm and $22\,$mm $ \times \; 22\,$mm, Menzel-Glaser). 
%Unlike soap films, this set-up is resistant against evaporation, fluid flow and rupture. Moreover, the absence of surfactants guarantees the natural swimming behaviour of cells.
}
\end{center}
\end{figure}

The experimental sample geometry is schematically represented in Fig.~\ref{fig:setup}.
Two glass cover-slips are  cleaned  with absolute ethanol on an ultra fine microfiber cloth before being exposed to oxygen plasma  and then dried with compressed air.
The main purpose of this treatment, which makes the coverslips hydrophilic, is to prevents bacterial cells and colloidal tracers from tethering to the glass.  % but it also facilitates loading of the samples.}
Next, we gently place 2~$\mu$l droplet of the suspension at the centre of the large cover-slip 
and locate the smaller cover-slip on top of it, thus creating a thin layer of bacterial suspension. The two cover-slips are then sealed  to prevent both evaporation and fluid flow (see figure \ref{fig:setup}).
The sample thickness was measured by focussing the microscope on colloidal or bacterial particles occasionally stuck on the top and bottom surfaces or on the top and bottom of the drop edges. 
%To measure the thickness of the experimental sample, we used bright field microscopy and focussed on colloidal or bacterial particles occasionally stuck on the top and bottom surfaces. 
This led to a lower bound of $\sim$ 5 $\um$. In view of the size of the colloidal particles, or bacterial cell body we estimate the actual thickness of the bacterial suspension to be in the range of $5-7$ $\um$.
%%%\subsection{Imaging}

\begin{figure}[h!]
\begin{center}
\includegraphics[width=7cm]{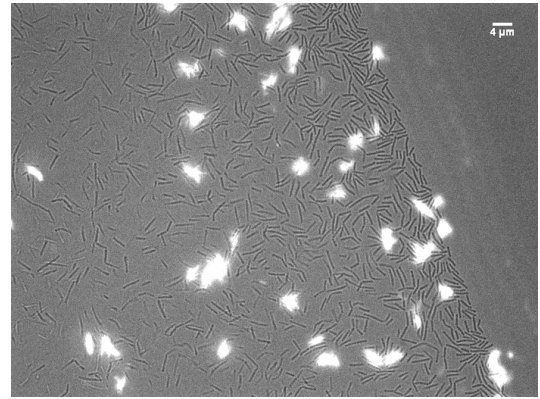}
\caption{\label{fig:snap0.09} Snapshot of a suspension of $\phi=0.09$ of \BS\
cells sandwiched between two oxygen-plasma-treated cover-slips, where $5\%$ of cells are fluorescently stained. Bacterial cell bodies appear black in the figure.}
\end{center}
\end{figure}

The suspension near the %oxygen contact line 
edge of the suspension drop 
was imaged using  an inverted microscope (TE2000-U, Nikon) with an oil immersion objective lens (Plan Flour $40\times$ OIL, NA 1.3, Nikon).  Images were captured using a cooled CCD camera (CoolSnapHQ$_2$, Photometrics) in conjunction with Metamorph software (Version: 7.5.5.0, Molecular Devices).
For fluorescent imaging, a GFP filter cube (49002 ET-GFP, Chroma Technology Corp) was used and the excitation light was synchronised with the camera in order to minimise exposure.
This fluorescence imaging provides high contrast images that allows for accurate particle tracking even if some background bright-field illumination is still present (see figure~\ref{fig:snap0.09}).
We recorded movies of the sample at 10 fps which were then analysed using a library of particle tracking routines written in IDL~\cite{tracking}.

%In all our experiments, we tracked the fluorescent beads for a minimum of 90 seconds.

%To track the motion of the \BS\ cells, we develop a novel experimental technique to stain the flagella of each cell, thus enabling us to follow all the trajectories of alive cells.

%At last, fluorescently labelled cells are diluted about 500-times before being used for measurements.
Fig.~\ref{fig:snap0.09} shows a still frame from a movie (excerpt available in supplementary material) of a suspension with about $\phi=0.09$
surface fraction (computed as the fraction between total bacterial area and sample area, by assuming that a single bacteria covers an area of $4 \um^2$), where
$ 5 \%$ of cells are fluorescently stained: in the image, 
it is even possible to identify the swimming modes of cells, where flagella 
dispersion indicates cell tumbling.
From this image it is also apparant that there is an accumulation of bacteria near the drop edge. {\BS} is an aerobic bacterium which moves towards the oxygen rich area near the egde, leading to an increase in bacterial density. However, a steady-state is established quickly ($\sim 1$ min) and by the time during data aquisition is started there is no more apparent change in bacteria density: 
%as there is an oxygen gradient in steady state in our sample. 
the clustering affects a region of about $20 \um$ close to the contact line which does not hamper our measurements as our field of view extends $\sim 100\um$ away from the interface.

%%% Figure 3 %%%

\section{Computer simulations} \label{sec:simulations}

We perform computer simulations on a similar system, i.e. a suspension of self-propelled particles 
with surface fraction of 0.03 and 0.09, respectively, containing few colloidal particles (about $1 \%$ of the total number of particles simulated). 
In our work we choose two independent ways to represent the self-propelled particles:  
a Brownian dynamics simulation of self-propelled dumbbells, and a generalisation of the two-dimensional vectorial Vicsek 
model~\cite{Vic1:1995/a}.
Both simulations incorporate passive colloids, which interact with the self-propelled particles via purely repulsive interactions.
Both self-propelled cells and colloids are chosen to have the same size ratio as in the experiments. 

In our experiments, colloids have a diameter of $\sigma_c \sim 4 \um$. In a bulk sample they were found to diffuse at about $D \sim 0.1\um^2/\text{s}$ (in agreement with Stokes-Einstein law),
whereas bacteria swim at a speed of  $v_0 \sim 30 \um/$s. One may therefore determine a dimensionless ``Peclet'' number as 
$Pe=\frac{v_0 \sigma_c}{D}$, which with our numbers is then $\sim 1100$. We note that the effective diffusion for a $4\um$ colloid in a thin sample may be smaller than the bulk value, by a factor of $\sim 0.5-0.6$ considering an effective thickness of $7 \um$~\cite{dufresne}. Considering the error and the variations in the numbers above (as well as the variation in bacterial cell size) it is more reasonable to consider a range of values for the Peclet number: comparison with experimental data suggests that  a range of $Pe \sim 700$ -- $1400$ is physically relevant.

\subsection{Self-propelled dumbbells}

The first avenue we propose to simulate a suspension of self-propelled particles with a small number of Brownian 
colloidal particles as tracers, is to use a system of ``active'' (in the sense of self-propelled) dumbbells. 

We equilibrate a quasi two-dimensional system of $N_d$=900  homomeric hard-dumbbells  in an {\it NVT} ensemble using cubic periodic boundary 
conditions in $x$-$y$ (the slab is 1 particle diameter height in the $z$ direction). Each dumbbell is made of two spheres with the same diameter $\sigma$, and each sphere interacts with the other $2 N_d-1$ 
by means of a truncated and shifted Lennard-Jones potential 
\begin{equation}\label{LJpot}
V_{LJ}(r) = 4 \epsilon \left[ \left( \frac{\sigma}{r}\right)^{12}-\left(\frac{\sigma}{r}\right)^6\right] - V_{LJ}(r=r_c)
\end{equation}
for $r\le r_c$ ($r=r_i-r_j$, with $i,j=1,...,2N_d$), whereas $ V_{LJ}(r)=0$ for  $r > r_c$, 
where $r_c=2^{1/6}\sigma$ and $\epsilon$ defines the strength of the interaction. 
Having chosen such short range cutoff (that corresponds to the minimum of the interaction potential),
 allows us to eliminate the attractive contribution of the potential and make the spheres ``effectively'' hard. 
Two spheres within a dumbbell are also ``glued'' together by means of a stiff harmonic 
potential $V_{H}(r)=K(r-r_0)^2$, where $r_0=\sigma$ is the equilibrium bond distance and $K=\frac{1}{2}k=10 \epsilon /\sigma^2$ (with $k$ the spring constant).

Simulations are carried out using the open source LAMMPS Molecular Dynamics package (see on-line documentation at http://lammps.sandia.gov/ and Ref.\cite{LAMPPS}).  
The equations of motion governing the dumbbell motion 
in an implicit solvent (hence neglecting hydrodynamic interactions) are given by the (under-damped) Langevin equations acting on each sphere:
\begin{equation}\label{eq:LD}
m \frac{d^2 \mathbf{r}_i}{dt^2} = - \zeta \frac{d \mathbf{r}_i}{dt} -  \frac{d V }{d \mathbf{r}_i} + \mathbf{F_r} + \mathbf{F_a}
\end{equation}
where $m$ is the mass of a sphere, $\zeta$ the friction coefficient ($\zeta=m \gamma$ with  damping coefficient $\gamma$), $V$ 
the total conservative potential acting on each particle ($V=V_{LJ}+V_H$) and $\mathbf{F_r}$ the force due to the solvent  at temperature
 $T$  -- following the fluctuation-dissipation theorem which we expect to hold in the $\mathbf{F_a}=0$ limit, we take this to be proportional to $\sqrt{k_B T \zeta }R(t)$ with $R(t)$ a 
stationary Gaussian noise with zero mean and variance $\langle R(t) R(t')  \rangle = \delta (t-t')$.
Self-propulsion is introduced via an extra force ($\mathbf{F_a}$) acting on each sphere such that the force 
 acting on the dumbbells has a modulus of $2 |\mathbf{F_a}|$ and the direction given by the dumbbell ``director'', 
which joins its front bead to its rear one (front and rear are chosen at the beginning of the simulation randomly). Such a director 
field corresponds then to the one used in polar nematic liquid crystals.

We then add to the suspension of self-propelling dumbbells  $2\%\, N_d$ of hard-sphere-like colloids, with $\sigma_c=2 \sigma$, 
and interacting with other constituents (passive or active) via the same truncated and shifted Lennard-Jones potential, where now 
$\sigma$ in eq.\ref{LJpot} is half the sum of the hard core diameter of the two interacting particles.

We start the simulations from an initial configuration where particles positions are random, while velocities come from a uniform random distribution 
at a given temperature of $T=0.001\epsilon/k_B$. In our simulations, we set  $\gamma=10 \tau^{-1}$ ($\tau$ is the time unit, 
$\sqrt{\frac{m \sigma^2}{\epsilon}}$, with mass set to one), 
the time step $\delta t = 10^{-3} \tau$, and typically evolve the system  for $10^6$ time steps. 
We choose the applied friction in order to simulate an effectively over-damped motion on the length scale 
of the particle size, as it is realistic for bacteria in our experiments. 
Parameters such as the temperature and the amplitude of the force  $F_0$, are chosen in order to lead to a Peclet number comparable 
to the one found in experiments ($Pe=1100$ when the average bacteria velocity is $30 \um/\text{s}$).
 More details on this model of active dumbbells and its properties will be presented elsewhere\cite{changeloda}.

Our simulations of the self-propelled dumbbells are for some aspects similar to 
those performed by Ortiz and coworkers \cite{ortiz}, who did not have a specific data set 
as target for their work, and used a minimal model of swimmers able to capture the far-field 
hydrodynamics, focusing on  collective dynamics in a confined suspensions. 
Our approach employs a purely Brownian dynamics simulations, therefore  by comparing 
with the trends presented in figure 2 of Ref.~\cite{ortiz} it is possible to single out the effect of hydrodynamic interactions.

In what follows, we will indicate with a subscript $b$ every variable that refers to the bacteria (or self-propelled particles), 
and with a subscript  $c$ every variable that refers to the colloids.
To quantitatively compare numerical results with experiments, we consider the diameter of the colloid $\sigma_c$ as unit 
length when measuring the mean square displacement of colloids, $\sigma$  as unit length when measuring the mean square displacement of
bacteria/self-propelled particles, 
and the Brownian time $t_0=\frac{\sigma_c^2}{D}$ (where $D$ is the diffusion coefficient of the colloids) as the unit time for both colloids and bacteria. 
To make contact with experiments, the surface fraction $\phi_b$ is set to $\phi_b=0.03$ and 
$\phi_b=0.09$, respectively.

Furthermore, we tune the value of the active particle velocity $v_0$ such that the dimensionless  number $Pe=\frac{v_0 \sigma_c}{D}$ is kept  constant and within the experimental range -- 
we loosely refer to this as the ``Peclet'' number even though one should note that this ratio is defined with the 
diffusion of the colloidal particles and the velocity of the self-propelled ones. Such a choice is justified 
as $D$ and $v_0$ give the natural diffusion and velocity scale in the model, as they are input to our algorithms, 
whereas for instance the diffusivity of bacteria can be found via numerical simulations.

\subsection{Vectorial Vicsek model}

As a simplified simulation to study the motion of bacteria as self-propelled particles, we also implement the two-dimensional Vicsek model. 
The Vicsek model was originally proposed to study flocking behaviours~\cite{Vic1:1995/a} as discussed in the original paper and later on e.g. 
in Ref.~\cite{chate4:2001/a}. Here, we use the Vicsek model to study bacterial suspensions, in view of the facts that (i) bacteria can be considered as 
self-propelled particles and (ii) they align with each other when they are close together, either via steric collisions or through hydrodynamic 
interactions~\cite{gareth}. The Vicsek model is an interesting example of an active non-equilibrium system, where activity
 leads (via self-propulsion)  to the existence of an order-disorder transition even in 2D, in violation of the Mermin-Wagner 
theorem which works for equilibrium systems.

Here we use the so-called ``vectorial'' Vicsek model, whose dynamical rules are as follows~\cite{chate4:2001/a,baglietto}:
the updated direction of the $i-$th particle is identified by an angle, $\theta_i(t+ \Delta t)$, which is computed by (i) 
first adding noise to the direction of each particle within a given distance ($r_0$) and (ii) averaging over these angles
\begin{equation}\label{eq:theta1}
\theta_i(t+ \Delta t) =  {\rm arg} \left [ \sum_{j=1}^{n_b^{\prime}} \exp^{i \theta_j (t)} + \varepsilon n_b^{\prime}  \exp^{i \xi_i(t)}  \right ]
\end{equation}
where ${\rm arg}$ denotes the argument of a complex number, and determines the direction of the velocity vector ${\mathbf v}_i$, $n_b^{\prime}$ is the 
number of neighbours within $r_0$ at time $t$ including particle $i$ itself, $\xi_i(t)$ is a delta correlated white noise ($\xi_i \in [-1,1]$) and $\varepsilon$ is the noise strength ($\varepsilon \in [0,1]$).

A relevant order parameter for the Vicsek model is the mean particle velocity ($\varphi = \frac{1}{N_b v_0} \mid \sum_{i=1}^{N_b} \mathbf{v}_i\mid$, 
with $v_0$ the modulus of the speed and  $N_b$ total number of Vicsek particles), as a function of noise ($\varepsilon$) and surface fraction ($\phi_b$).

Here we generalise the Vicsek model to incorporate a small number of colloidal particles (a similar simulation method was proposed in Ref.~\cite{chate4:2001/a}, 
but no attempt was made there to get quantitative predictions of mean square displacements). In our implementation, we introduce a finite 
size for {\em both} Vicsek particle (which now have diameter $\sigma_b$ as opposed to being point-like) and passive colloids (with diameter $\sigma_c$).

In our algorithm, at every time step $\Delta t$ each Vicsek particle evolves according to 
\begin{eqnarray}\label{positionvic}
{\bf r}_{i,b}(t + \Delta t) & = & {\bf r}_{i,b}(t) + {\bf v}_{i,b}(t + \Delta t)\Delta t       \\ \nonumber
& & - \frac{\Delta t}{\zeta_{b}}\sum_{j=1}^{N_c}\nabla U_{bc}(r_{ij})  -  \frac{\Delta t}{\zeta_{b}}\sum_{l=1}^{N_b}\nabla U_{bb}(r_{il})
\end{eqnarray}
where ${\bf r}_{i,b}$ is the particle displacement, ${\bf v}_{i,b}$ is the particle velocity, $\zeta_{b}=3 \pi \eta \sigma_b$ is 
the particle drag ($\eta$ is the viscosity) and $N_c$ is the number of colloids; $U_{bc}$ 
is a soft repulsive potential between the Vicsek particles and the colloids, whose gradient is constant and different from zero
 only when a particle and a colloid overlap, whereas $U_{bb}$ is a similar repulsive potential between colliding Vicsek particles. 
We choose $\frac{|\nabla U_{bc}|}{3 \pi \eta \frac{(\sigma_b+\sigma_c)}{2}}=\frac{|\nabla U_{bb}|}{3\pi \eta \sigma_b} \simeq 0.37$. Furthermore, 
choosing $v_0 \Delta t < r_0$ guarantees that particles have enough time to interact with  their neighbours.

Passive colloids are modelled as two dimensional disks with diameter $\sigma_c$  evolving via a similar over-damped Langevin dynamics as in eq.\ref{positionvic}, given by
\begin{eqnarray}
{\bf r}_{i,c}(t + \Delta t) & = & {\bf r}_{i,c}(t) 
- \frac{\Delta t}{\zeta_{c}}\sum_{j=1}^{N_b}\nabla U_{bc}(r_{ij})  \\ \nonumber
&- & \frac{\Delta t}{\zeta_{c}}\sum_{l=1}^{N_c}\nabla U_{cc}(r_{il})
+ {\mathbf{F_r}}
\end{eqnarray}
where ${\bf r}_{i,b}$ is the colloidal displacement, $\zeta_c=3 \pi \eta \sigma_c$, $\mathbf{F_r}$ is a random
noise associated to the colloidal diffusion coefficient 
as in Eq.\ref{eq:LD}, and 
$U_{cc}$ is the repulsive potential between colliding colloids, ($\frac{|\nabla U_{cc}|}{3\pi \eta \sigma_b} \simeq 0.37$) and 
$U_{bc}$ is the same as before. 
We simulate a two-dimensional square box with size $L$ and periodic boundaries, setting $N_b=1000$ and $N_c=10$ and 
choose $\sigma_c=2\sigma_b$.

A crucial parameter within the Vicsek model is the amount of local ordering in the system, $\varphi$, which depends on a combination of the noise amplitude ($\varepsilon$)
the interaction range ($r_0$) and $v_0$. When $\varphi$ is large, Vicsek particles form large flocks and  colloids move 
ballistically in the direction of the flock; at intermediate $\varphi$, Vicsek particles present low order and 
form short-lived flocks and colloids present a crossover (identified by a crossover time $t_c$) from superdiffusive 
to diffusive behaviour; for small $\varphi$, Vicsek particles are randomly oriented  and colloids simply diffuse, 
albeit faster than purely Brownian particles. 
In our simulations, we choose an intermediate noise amplitude, $\varepsilon=0.55$. This choice 
is motivated by the experimental observation that the suspension does not contain large flocks of bacteria, and that experiments 
show superdiffusive behaviour at early times.
Tuning $r_0$ also affects the behaviour (at a given  $\varphi$) and the position of the crossover time $t_c$: the smaller $r_0$, 
the earlier $t_c$ occurs. In our simulations, we observe the best (qualitative, see below) match with the experimental 
trends with $r_0=2\sigma_c$ (even though we also tried $r_0=\sigma_c$ and $r_0=3\sigma_c$): 
our best results correspond to a value of the ``Peclet'' number  ($Pe=\frac{v_0 \sigma_c}{D}$) of the order of $\sim 1400$ to which we stick to in the following.

\section{Results} \label{sec:results}

\begin{figure}[h!]
\begin{center}
\includegraphics[width=6.5cm]{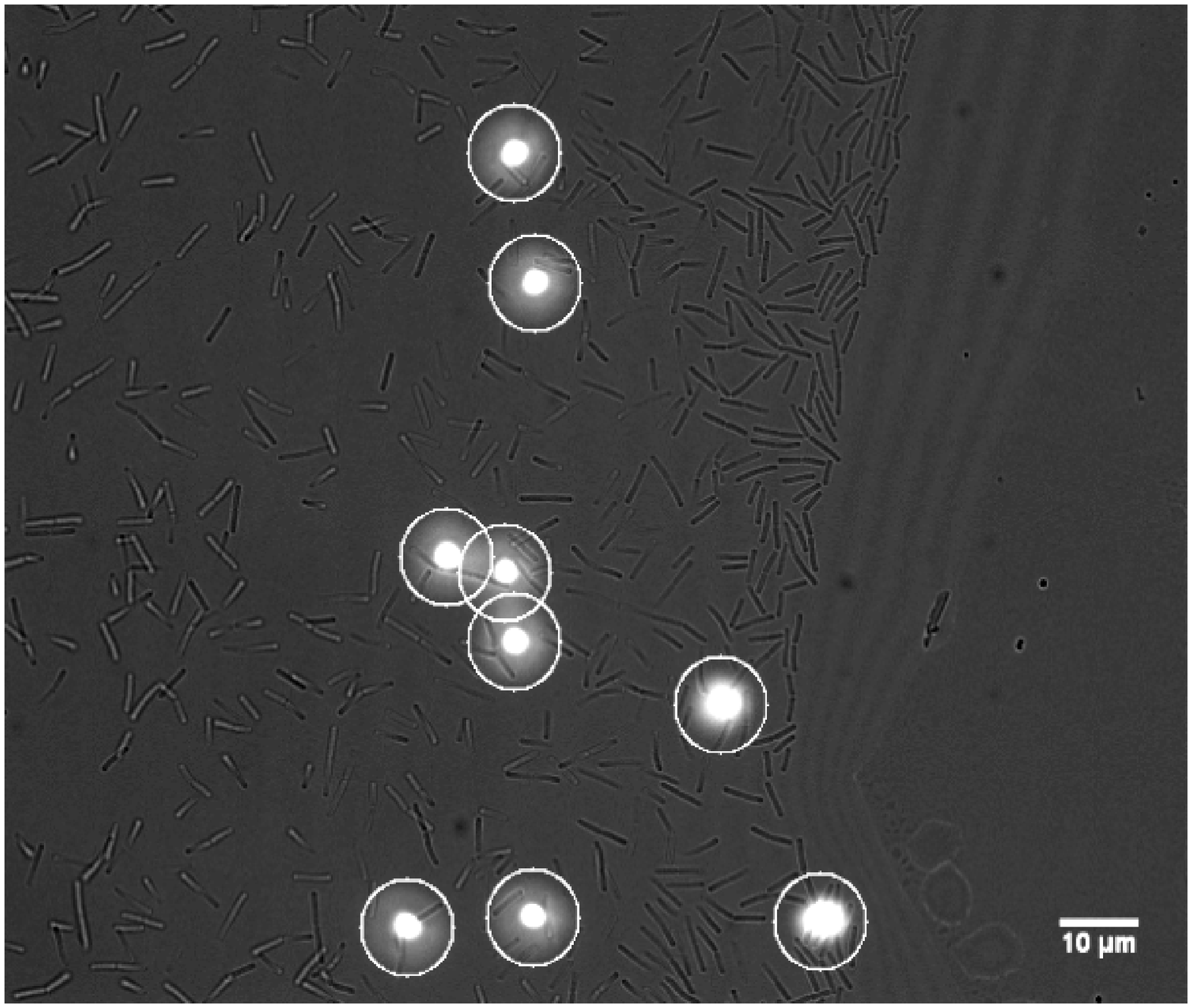}\\
\includegraphics[width=6.5cm]{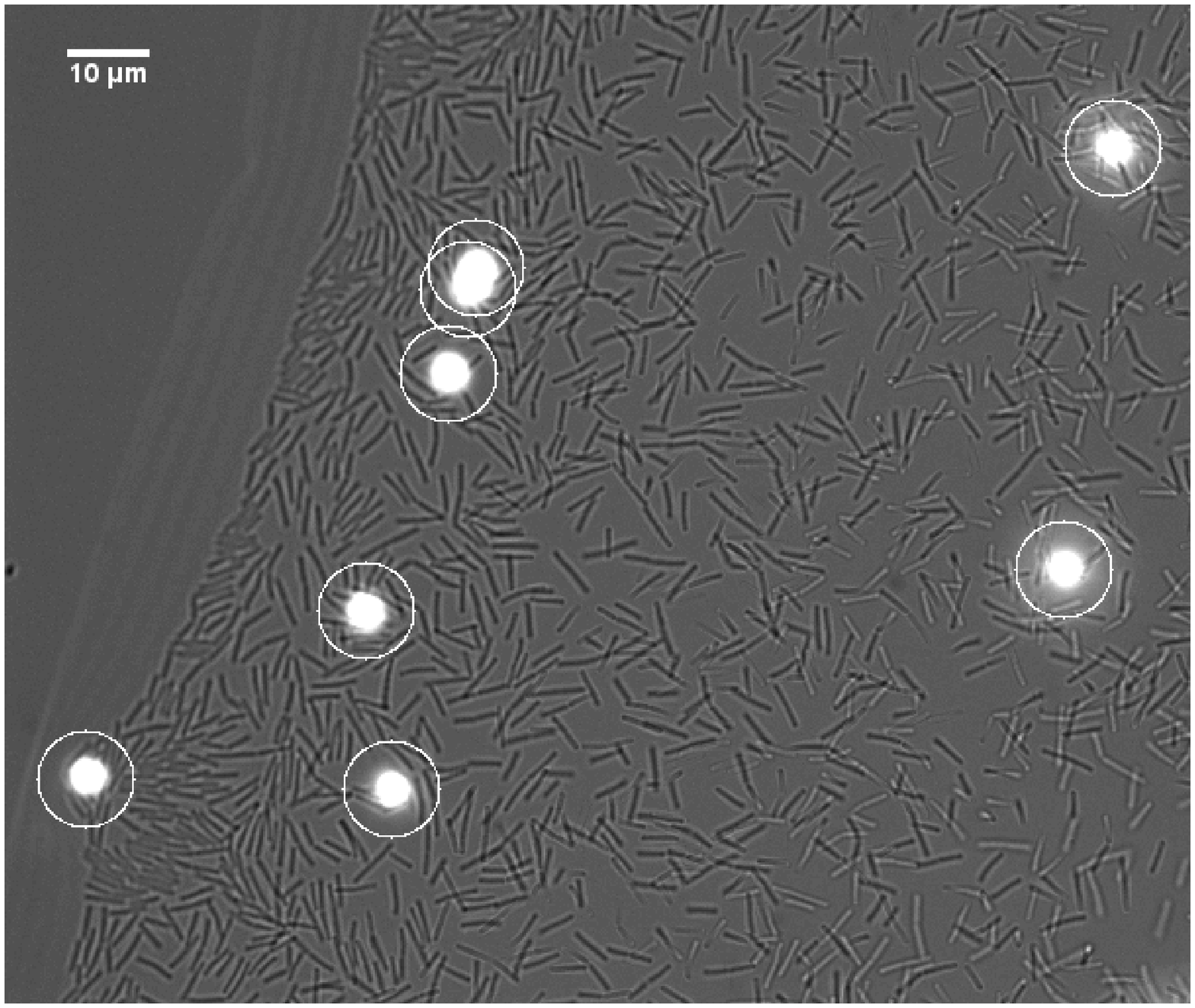}
\caption{ (a) \BS\ cells suspended between treated cover-slips at 
$\phi_b = 0.03$ (top) and  $\phi_b=0.09$ (bottom), and $4 \mu m$ fluorescent beads.\label{fig:snapexpe}}
\end{center}
\end{figure}

To start our analysis, we show some typical snapshots of the bacterial suspension we study. Figure~\ref{fig:snapexpe} 
shows some optical microscope images showing both bacteria and fluorescent beads, at the two densities we have recorded 
our data for, $\phi_b=0.03$ (top) and $\phi_b=0.09$ (bottom). It can be seen that in both cases there are significant 
inhomogeneities in the bacterial density. Our experiments also suggest that there is collective dynamics of 
bacteria at both the densities studied (see Supplementary Movie 1 for a video of the dynamics of bacteria 
at $\phi_b=0.09$). Such a collective behaviour may be due to either collisions or 
hydrodynamic interactions between the bacteria. Our experiments and simulations however both suggest there is negligible nematic ordering in the sample at these densities.

\begin{figure}[h!]
\begin{center}
\includegraphics[width=4.5cm]{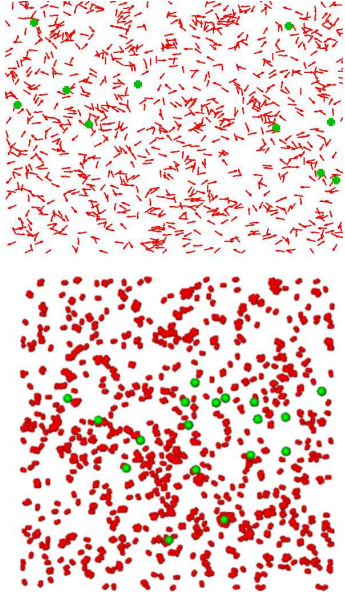}
\end{center}
\caption{\label{fig:snapsimu} Top:  Bacteria particles simulated using the vectorial Vicsek model (red arrows) at surface fraction $\phi_b = 0.09$. Passive beads are represented by green dots. Bottom: Bacteria simulated as self-propelled dumbbells (in red) and passive tracers (in green)  at the same surface fraction.}
\end{figure}

Figure~\ref{fig:snapsimu}  shows snapshots taken from our numerical simulations
 (where bacteria are either self-propelled dumbbells (red dumbbells in the bottom panel) 
or Vicsek particles (red arrows in the top panel). 
These snapshots also give evidence of a significantly inhomogeneous density profile which qualitatively matches the experimental one. 

In order to explore more systematically and quantitatively the dynamics of both colloidal 
particles and bacteria in experiments and simulations, we recorded time series of the 
mean square displacement (MSD) of both -- the latter is defined via $ MSD_{b,c} (t) = \langle \left[ \vec{r_i} (t) - \vec{r_i}(0) \right]^2\rangle_{b,c}$, 
where the average runs over the total number of bacteria or colloids (the MSD computed from numerical 
simulations is then averaged over tens of realisations).
The mean square displacement depends on time as $t^\alpha$, where $\alpha = 1$ when particles move 
diffusively, $ 1< \alpha < 2$ when they move superdiffusively and $\alpha = 2$ when they are in a ballistic regime.

\begin{figure}[h!]
\begin{center}
\includegraphics[width=7cm,clip=]{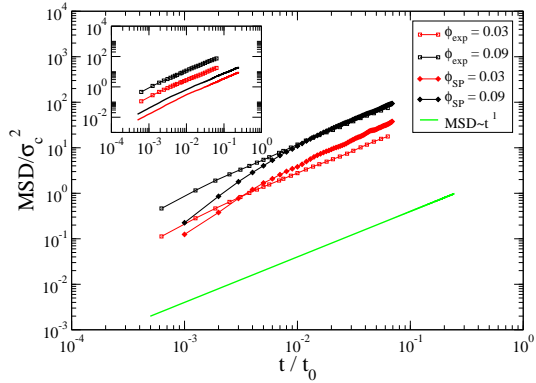}
\caption{\label{fig:msd}Beads MSD (in units of $\sigma_c$)
as a function of time (scaled with $t_0=\sigma_c^2/D_c$) at  
$\phi_b=0.03$ (red curves) and $\phi_b=0.09$ (black curves). 
Open squares are experiments, filled diamonds results from simulations with SP-dumbbells.
The thick green line shows the Stokes law prediction for the
MSD for diluted Brownian colloidal particles in
the absence of any swimmers (this agrees well with our data for isolated colloids in solutions, see above).
Inset: Beads MSD  from Vicsek simulations and experiments (both in units of $\sigma_c$)
as a function of time (scaled with $t_0$) at  the same $\phi_b$
($\phi_b=0.03$ (red curve) and $\phi_b=0.09$ (black curve)). 
Open squares are the experimental results (same as the ones in the figure).}
\end{center}
\end{figure}

We first analyse the mobility and dynamics of the colloidal tracers. 
Figure \ref{fig:msd} shows the mean square displacement of the colloids as a 
function of time at $\phi_b=0.03$ and $\phi_b=0.09$ estimated from experiments 
and simulations where the self-propelling particles where 
either the self-propelled dumbbells ($SP$) or the vectorial Vicsek particles (in the inset).

>From Fig.~\ref{fig:msd}, it is apparent that the MSD computed from the SP dumbbells 
matches the experiments better than the MSD from the Vicsek particles -- the latter 
largely underestimates the passive colloidal mobility. 
The quantitative discrepancy  between our two simulations may be either due to the simplified steric representation of colloids and 
bacteria in the Vicsek model (which are both spherical) or to the presence of residual flocking, which 
may lead to unrealistic density inhomogeneities in the sample, possibly leading to an underestimation 
of the collisions and of the interactions with the diffusing colloids. 
Nevertheless, we find that the trends of the curves obtained with the Vicsek simulations are qualitatively 
in agreement with the ones measured both experimentally and with the SP dumbbells simulations: 
when plotting the MSD of the colloids as a function of time, we observe a transition from a 
short time regime where colloids move in a superdiffusive way, to a long time regime where colloids  
move diffusively.  
Moreover, we observe that colloids move much faster 
(about 2 order of magnitude) than their purely Brownian counterpart ($MSD=4tD$ shown as the
green line in Fig.~\ref{fig:msd}).

The crossover times predicted by our simulations (both with dumbbells and Vicsek particles) 
together with those found experimentally are summarised in
Table~\ref{tab:experiment}, where we also include the early and late time exponents characterising the
MSD. 
% Table %
\begin{table}[h!]
\begin{center}
\caption{\label{tab:experiment} Results from experiments [Exp] and numerical simulations 
for colloids in self-propelling dumbbells [SP] (at different $Pe$ number) and in Vicsek particles [Vic]
at $\phi_b=0.03$ and $\phi_b=0.09$:  the crossover time $t_c$ is expressed in 
units of $t_0=\sigma_c^2/D_c$ and $\alpha_{s}$ and $\alpha_{l}$ correspond to the exponent of $t^\alpha$ for short and long time, respectively. 
The last column shows the value of the MSD scaled by $\sigma^2$ at the crossover, ${\rm MSD}(t_c)/\sigma_c^2$. Relative errors in the experimental and simulated  estimates of the crossover times are about $10\%$. Typical errors on exponents do not exceed $\pm 0.1$.}
\begin{tabular}{lcccccc}\hline\hline
                         &
$\phi_b$ & $t_{c}/t_0$  & $\alpha_{s}$                & $\alpha_{l}$ & $Pe$ & ${\rm MSD}(t_c)/\sigma_c^2$        \\ \hline
$\mbox{Exp}$ &
$0.03   $ &
$0.0042 $ &
$1.2    $ &
$1.0    $ &
$[700-1400]     $ &
$1.1$ \\
$\mbox{Exp}$ &
$0.09   $ &
$0.0055 $ &
$1.2    $ &
$1.0    $ & 
$[700-1400]     $ &
$6.0$ \\  \hline\hline

$\mbox{SP}$ &
$0.03   $ &
$0.0050 $ &
$1.5    $ &
$1.0    $ &
$800     $ &
$0.7$   \\
$\mbox{SP}$ &
$0.09   $ &
$0.0060 $ &
$1.9    $ &
$1.0   $ &
$800     $ &
$3.7$ \\  \hline\hline

$\mbox{SP}$ &
$0.03   $ &
$0.0045 $ &
$1.6    $ &
$1.0    $ &
$1100     $ &
$1.7$   \\
$\mbox{SP}$ &
$0.09   $ &
$0.0056 $ &
$1.8    $ &
$1.0    $ &
$1100     $ &
$5.3$\\  \hline\hline

$\mbox{Vic}$ &
$0.03   $ &
$0.0040 $ &
$1.4    $ &
$1.0    $ &
$1400     $ &
$0.1$   \\
$\mbox{Vic}$ &
$0.09   $ &
$0.0050 $ &
$1.4    $ &
$1.0    $ &
$1400   $ &
$0.3$   \\  \hline\hline
\end{tabular}
\end{center}
\end{table}

Note that in experiments the Peclet number is in the range $[700-1400]$, 
so that simulations with SP dumbbells are done with Peclet number of 800 and 1100 (simulations with the Vicsek 
particles are shown for a Peclet number of 1400 which yields the best results).
As shown in Table~\ref{tab:experiment}, there is a good quantitative agreement between the 
estimates of the crossover time, at both bacteria surface fractions, obtained from experiment and 
simulations. Even though the crossover time is well predicted, the simulations at all Peclet number slightly overestimate the 
exponent associated with the superdiffusive regime. Quite possibly, the discrepancy may be due to our
simplified description of the swimming mechanism of \BS~(see Discussion below).
Our simulation data also suggest that the crossover time decreases when 
the Peclet number increases, at least in the range we considered. 
\if{We interpret this result as due to the fact that at higher propulsion velocity, there is 
a larger number of collisions between colloidal particles and SP ones, and this anticipates the onset of the diffusive regime.}\fi
All the data presented in Table~\ref{tab:experiment}, both experimental and numerical, also show that the crossover time 
increases with the density of the system (although the experiments alone would not be conclusive as there is only a 
small change in the MSD exponent). In what follows, we will run the simulations only for the system 
containing dumbbells, and only at the Peclet number corresponding to the average bacteria velocity measured in the 
experiments (i.e.$30 \um/\text{s}$). 

In order to investigate more in depth the dependence of the crossover time with 
density, we simulate several systems containing SP dumbbells and colloids, 
and systematically vary the dumbbell surface fraction, $\phi_b$. 
Our results for the colloids are presented in Table~\ref{tab:simulation1}
(data of $t_c$, $\alpha_{s}$ and $\alpha_{l}$ are obtained plotting 
the MSD of colloids).
\begin{table}[h!]
\begin{center}
\caption{\label{tab:simulation1} Results for $t_c$ (in units of [$\sigma_c^{2}/D_c$]), $\alpha_{s}$ and $\alpha_{l}$ 
for colloids immersed in suspensions with increasing $\phi_b$. Relative errors in the simulated  estimates of the crossover times are about $10\%$.}
\begin{tabular}{lcccc}\hline\hline                         &
$\phi_b$ & $t_{c}/t_0$ &  $\alpha_{s}$  &  $\alpha_{l}$          \\ \hline

$\mbox{}$ &
$ 0.001  $ &
$ 0.010    $ &
$ 1.7   $ &
$ 1.0    $ 
\\

$\mbox{}$ &
$ 0.01  $ &
$ 0.0065    $ &
$ 1.7   $ &
$ 1.0    $ 
\\

$\mbox{}$ &
$ 0.03  $ &
$ 0.0045    $ &
$ 1.6   $ &
$ 1.0    $ 

\\
$\mbox{}$ &
$ 0.06  $ &
$ 0.0050    $ &
$ 1.8   $ &
$ 1.0    $  
\\
$\mbox{}$ &
$ 0.09  $ &
$ 0.0056    $ &
$ 1.8   $ &
$ 1.0    $ 
 \\
$\mbox{}$ &
$ 0.12  $ &
$ 0.0065    $ &
$ 1.9   $ &
$ 1.0    $  
\\
$\mbox{}$ &
$ 0.15  $ &
$ 0.0075    $ &
$ 1.8   $ &
$ 1.0    $ 
 \\
\hline\hline
\end{tabular}
\end{center}
\end{table}

The data presented in the table give evidence of an intriguing non-monotonic relation between 
$t_c$ and $\phi_b$. When the system is diluted ($0.001 \le \phi_b \le 0.03$), colloids diffuse 
around and sometimes are pushed by the few swimming dumbbells in the system: in this range a higher density
gives rise to more randomised collisions, hence the crossover to diffusion occurs earlier. 
However, for $\phi_b\ge 0.03$ this trend reverses and an increase in density leads to an increase in
the crossover time. 

\begin{figure}[h!]
\begin{center}
\includegraphics[width=7cm,clip=]{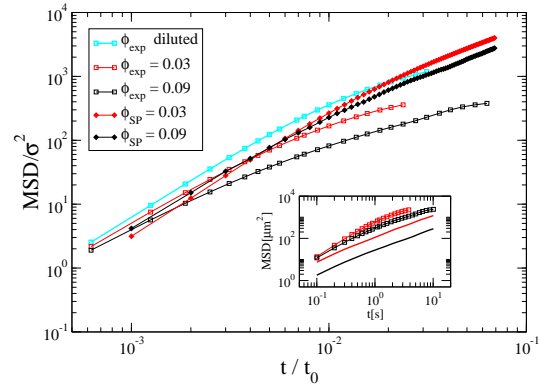}
\caption{MSD  of bacteria/self-propelled particles (in units of $\sigma$)
as a function of time (scaled with the Brownian colloidal time $t_0$)
at two bacterial surface fractions, $\phi_b=0.03$ (red curves) and $\phi_b=0.09$ (black curves).
The cyan curve represents the experimental results for a diluted system. 
Open squares are experimental results, filled diamonds are simulations with SP dumbbells. 
In the inset: beads MSD (from Fig.\ref{fig:msd}) and bacteria  MSD (MSD is measured in $\um^2$
and $t$ in seconds).
\label{fig:beadspart}}
\end{center}
\end{figure}

To further understand the dynamics of our system, and to understand the mechanism leading to
the non-monotonic trend of the crossover time with density, we set out to study more in detail the motion of the bacteria, or SP dumbbells. 
To this aim, we have measured, by means of both experiments and computer simulations, the mean square displacement of 
bacteria/SP particles in bacterial bath at different surface fraction ($\phi_b=0.03$ and $\phi_b=0.09$)
as well as in the dilute limit (in which bacteria are so dilute that they do not interact with each other).
The results are shown in Fig.~\ref{fig:beadspart}.

As shown in the inset (for experiments only, simulations show the same trend), we also observe that bacteria (open squares)
move much faster than the beads (thick continuous lines) at the same surface fraction ($\phi_b=0.03$ in red and 
 $\phi_b=0.09$ in black).
At short time scales, the data coming from simulations of the SP dumbbells agree 
well with the experimental ones. 
Qualitatively, the dumbbell simulations and the experiment also agree on the trend that the higher the 
density becomes, the slower the long-time diffusion of the bacteria/self-propelled particles is. 
This effect comes from crowding due to the steric effects of particles. 
Quantitatively, on the other hand, the SP particles move 
faster than the bacteria in the experiments. This may be due to additional hydrodynamic drag felt 
by the bacteria in the experiments, as they are swimming in a thin fluid layer
(thickness $\sim 5-7 \um$) sandwiched between two cover-slips, so that wall effects may play a role.

\begin{figure}[h!]
\begin{center}
\includegraphics[width=7cm,clip=]{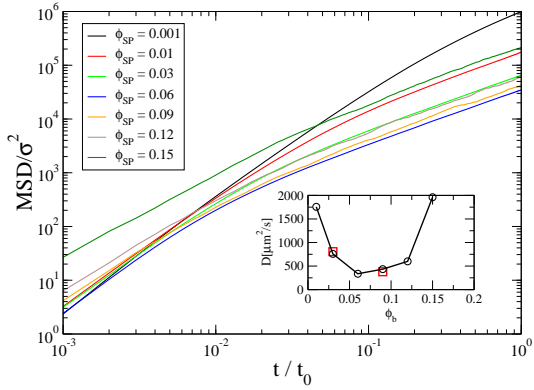}
\caption{MSD (in units of $\sigma$) of SP dumbbells as a function  of time (scaled with the Brownian colloidal time $t_0$)
at various surface fractions. The inset shows the long-time diffusion (in units of $\mu m^2$/s) of 
the dumbbells as a function of $\phi_b$ (the red 
squares are the experimental diffusion coefficients multiplied by a factor of 7 -- this shows that in spite of the quantitative
discrepancy the trend is the same).
\label{MSDbugs-phi}}
\end{center}
\end{figure}

The result that the MSD of bacteria at long time scales decreases with increasing density, 
seems at first sight incompatible with the results reported in Ref.~\cite{goldstein1} which shows that, due to cooperative effects, dense suspensions 
of {\em B. subtilis} swim collectively {\em faster} than single bacteria. One possible explanations is that our densities 
are too dilute in order for collective effects to play a role. But what is the mechanism leading to the onset of 
cooperativity? Does this arise from hydrodynamic interactions? Or is the steric exclusion enough for it to appear? 
In order to explore this possibility, we have simulated  significantly denser dumbbell suspensions which we were not 
able to study experimentally. The resulting MSD plots for the self-propelled particles are shown in Fig.~\ref{MSDbugs-phi}.

>From the MSD data in Fig.~\ref{MSDbugs-phi}, one can estimate the (late time) diffusion coefficients of the dumbbells as a function of $\phi_b$. 
These are plotted in the inset of Fig.~\ref{MSDbugs-phi}, where we observe again a non-monotonic trend with $\phi_b$, this
time of the SP particle diffusivity. First, as observed above, mobility and effective long time diffusion decrease with density, as it is increased from $0.03$ to $0.06$. After this, the long-time diffusion increases, and this enhancement becomes more and more dramatic with increasing density: for instance the MSD for $\phi_b=0.15$ is already an order of magnitude larger than the one at $\phi_b=0.03$. 
Therefore our simulations suggest that even steric interactions alone can lead, through over-damped collision, to long range correlations and an enhanced mobility. What is particularly interesting is that this effect is already well developed at a relatively modest density of $0.15$. This steric mediated cooperative motion is thus going to compete to the hydrodynamic cooperative motion, which is neglected in our Brownian dynamics simulations.

Fig.~\ref{MSDbugs-phi} demonstrates that the interplay between steric interactions (crowding) and (purely collisional) collective behaviour can give rise to a non-monotonic SP particle diffusivity  with density. We believe that the same interplay is at the base of the non-monotonic trend of the crossover time for colloidal particles as a function of $\phi_b$. To further consolidate this argument, we show in Table~\ref{tab:simulation2} the crossover time in the dynamics of the SP particles: once more, this quantity behaves non-monotonically with $\phi_b$. Table~\ref{tab:simulation2} shows the early and late MSD exponents, $\alpha_{s}$ and $\alpha_{l}$, which we found from the MSD of SP dumbbell in suspensions at various $\phi_b$. 
\begin{table}[h!]
\begin{center}
\caption{\label{tab:simulation2} Results for $t_c$ (in units of [$\sigma_c^{2}/D_c$]), $\alpha_{s}$ and $\alpha_{l}$ 
for self-propelling dumbbells in  suspensions with increasing $\phi_b$. Relative errors in the simulated  estimates of the crossover times are about $10\%$.}
\begin{tabular}{lcccc}\hline\hline                         &
$\phi_b$ & $t_{c}^{SP}/t_0$ &  $\alpha^{SP}_{s}$  &  $\alpha^{SP}_{l}$      \\ \hline

$\mbox{}$ &
$ 0.001  $ &
$ 0.090    $ &
$ 2.0   $ &
$ 1.5    $  \\

$\mbox{}$ &
$ 0.01  $ &
$ 0.040    $ &
$ 1.9   $ &
$ 1.1    $  \\

$\mbox{}$ &
$ 0.03  $ &
$ 0.030    $ &
$ 1.8   $ &
$ 1.0    $

\\
$\mbox{}$ &
$ 0.06  $ &
$ 0.020    $ &
$ 1.7   $ &
$ 1.0   $
\\
$\mbox{}$ &
$ 0.09  $ &
$ 0.025    $ &
$ 1.6   $ &
$ 1.0    $
 \\
$\mbox{}$ &
$ 0.12  $ &
$ 0.030    $ &
$ 1.5   $ &
$ 1.0    $
\\
$\mbox{}$ &
$ 0.15  $ &
$ 0.050    $ &
$ 1.5   $ &
$ 1.0    $
 \\
\hline\hline
\end{tabular}
\end{center}
\end{table}

We observe that while in the most dilute case the dumbbells initially move ballistically, the early time behaviour crosses over to superdiffusive for increasing $\phi_b$. Note that for the most diluted system ($\phi_b=0.001$), the  long-time exponent $\alpha_l$ indicates a superdiffusive behaviour, as the self-propelling dumbbells so rarely collide with their neighbours -- however for all higher surface fraction we reproduce the late time diffusive behaviour observed in experiments.

\section{Discussions and conclusions} \label{sec:discussion}

One of our main findings is that in both experiments and numerical simulations we observe that, at short times, passive colloids are dragged by bacteria and move in a super-diffusive, 
rather than ballistic, whereas at later times 
they slow down and diffuse, although their long-time diffusion coefficient is very different from, and larger than, their infinite dilution limit. The time at which the crossover occurs is a small fraction of the Brownian time 
which it takes for a particle to diffuse its own diameter, and its numerical value can be reproduced by our simulations with high accuracy.
The presence of a cross-over from super-diffusive to diffusive behaviour is in agreement with Wu's experiments of a two-dimensional 
diluted bacterial suspension on a soap film\cite{Wu:2000/a}, where at short time scales the system turned out to be super-diffusive 
with $\langle  \Delta r^2 (t) \rangle \sim t^\alpha$ and $1.5 < \alpha < 2$, slightly larger than in our experiments and closer to 
our numerical estimates~\cite{Wu:2000/a}. 

%%%Differences between Wu and Libchabers experiment:

Even though superficially our experimental work is similar to the one by Wu and Libchaber\cite{Wu:2000/a}, there are several important differences.
Firstly, the previous work used a different bacterium, {\it E. coli}, which is smaller 
(typically  $2 \times 1 \um$) than the {\BS} used here. This makes the bacteria in our experiment more similar in size to the passive 
tracer beads and, in combination with the thinner sample geometry, also ensures that the movement of our {\BS} cells is 
almost strictly {\em two-dimensional} as cells cannot tumble in the direction perpendicular to the cover-slips. 
Furthermore, as shown in Fig. 1, the bundle arrangement of flagella while swimming is quite different from the one in {\it E. coli}, which will be discussed in a separate manuscript~\cite{martin}.
Secondly, the previous work employed a soap film:  soap is a chemical repellent which 
can disrupt swimming of bacteria and furthermore bacteria and beads can get stuck at the air-water interface of the  film.
Thin soap films are also prone to drift or rupture due to evaporation and capillary forces. 
In our experiments the motile 
bacteria sample is instead suspended between two oxygen plasma treated cover-slips which makes them extremely robust 
against evaporation and rupture. The oxygen plasma treatment also makes the glass surfaces very hydrophilic and thus prevents cells and beads tethering to the surface.
The cover slips make it possible to use high resolution immersion objectives, improving particle image analysis 
hence overall accuracy in the determination of trajectories and mean square displacements. Therefore, our setup is 
particularly well suited for quantitative comparison with computer simulations. 
However, the two nearby glass surfaces also introduce non-slip boundaries which may lead to hydrodynamic slow-down 
of the motion, as the plates are just about $5-7 \um$ apart. This has not been taken into account in our simulations (or interpretation of the experiments). 
The experiments presented here rely on 
accurate tracking of the motion of individual particles.
Even with the high contrast images obtained with fluorescence microscopy this is time-consuming and it is difficult 
to monitor enough tracers over a sufficiently long time interval to obtain good statistics.
It has been recently demonstrated that  better statistics can be achieved by means of 
novel experimental techniques such as Differential Dynamic Microscopy~\cite{vincent}.

What is the superdiffusion-diffusion crossover physically due to?
In their work, Wu and Libchaber concluded that its physical origin could be attributed to the collective dynamics of bacteria.
But what ultimately leads to this collective behaviour? 
The latter might either be due to bacterial collisions, or to non-steric and hydrodynamic mediated interactions. While our 
dumbbell simulations faithfully consider the steric and geometrical properties of bacteria and colloids as well as their drag, 
long-range hydrodynamic interactions are neglected. Still, Fig. 8 clearly demonstrates that steric interactions alone (i.e. collisions) 
are able to lead to a collective behaviour, and also to the speed up of self-propelled particles at large densities, which 
was observed in concentrated bacterial baths, and was commonly attributed to hydrodynamics alone.

An important conclusion of our work is therefore that steric interactions via collisions play an important role, 
such that the collective behaviour is likely to come from a combination of this effect {\em and} hydrodynamic interactions. 
An accurate model of  hydrodynamic interactions in self-propelled particles is not a trivial issue, 
as accurate Stokesian dynamics simulations slow down dramatically with number of particles~\cite{pedley,pedley2}, 
and a more viable alternative might be the one suggested by Ref.~\cite{ortiz}. 
Our algorithm employing Brownian dumbbells is computationally less expensive and still 
leads to a good, semi-quantitative, agreement with experiments, which in many cases may be sufficient. 

Another interesting result from our analysis is that the Vicsek model, although qualitatively leading to the existence 
of a superdiffusive-diffusive crossover and to an enhancement of the Brownian diffusion of colloidal tracers, shows 
a poor quantitative agreement with the experiments. To some extent, this is quite surprising: while it is true that the Vicsek 
model does not resolve the rod-like shape of bacteria, it also does take into account the bacterial local alignment (which might be viewed as a rough 
way to mimic e.g. hydrodynamic effects) and has additional parameters which can control the local ordering of self-propelled particles -- 
in dumbbells this only comes from geometry and steric effects. In view of the performance of the Vicsek model shown in Fig.6, 
the comparison between dumbbell simulations and experiments becomes remarkable, given that there are essentially
 no free fitting parameters in the simulations (except the Peclet number which experiments only constraint to be in a range of realistic values).

While the MSD of the colloidal particles coming from the simulations of the suspension of SP dumbbells and colloids can be 
quantitatively compared with the experimental ones, the MSD of the self-propelled dumbbells are only in order-of-magnitude agreement
with the experimental bacterial ones. What is the reason for this behaviour?
One possibility is that in our simulations we did not take into account bacterial tumbling, which may 
be thought to affect more the bacteria than the colloidal particles. However, it would not be trivial to modify the simulations 
to more faithfully reproduce the swimming mechanism of \BS~ as this, to date, remains much less studied than that of {\it E. coli} -- 
for instance, bacterial flagella bundle differently, as suggested by our experiments  (Fig. 1 and Ref.~\cite{martin}). 
Another  possibility is that the neglect of hydrodynamic interactions is responsible for this quantitative discrepancy. Hydrodynamic interactions 
are thought to predominantly come through the forces exerted by the bacteria on the fluid. These
are not included in our treatment, whereas they may lead to an enhanced local flow. However a careful incorporation of the effect of such forces on the fluid may also not be trivial as it has recently been shown experimentally in~\cite{goldstein2010} that the dipolar (stresslet) contribution is in reality difficult to measure, and masked by near-field quadrupolar interactions (and far-field monopole contributions due to the fact that swimmers in practice are not neutrally buoyant).

Furthermore, for our experiments, due to the small thickness of the sample used experimentally, hydrodynamic interactions due to the flow field generated by the bacteria will be screened due to the presence of the nearby walls -- this may partially account for the good level of quantitative agreement with our simulations.

We should finally stress that it is quite possible that 
the effective density estimated in experiments is higher than the nominal one, due to inhomogeneities, which typically lead to 
a higher concentration at the edge of the droplet within which bacteria are confined. If the density were effectively larger, 
so would be the collective behaviour and the mean square displacement.

%Conclusions go down here
In conclusion, we have performed a joint experimental and numerical study of the behaviour of colloidal tracers in a bacterial 
fluid. With respect to previous simulation work, we have explicitly taken into consideration the finite size of both self-propelled 
particles (modelling bacteria), and of colloidal particles. 
The experimental work instead differs from previous work as we used \BS\ films and a novel experimental setup.
In agreement with earlier work, our experiments show that the effective diffusion coefficient of the colloidal particles is about
 two orders of magnitude larger than the Stokesian value for $4\um$ diameter spheres. By focusing on 
the traces of swimming bacteria and self-propelling particles, we show unambiguously that the early regime is superdiffusive, and we find that the 
crossover to diffusive regime occurs at a time which increases with bacterial cell density.
We have compared experimental determinations of the crossover times with estimates from computer simulations, where the bacteria 
were modelled either via self-propelled (SP) dumbbells or via the (vectorial) Vicseck model. This is 
the first quantitative comparative study of this kind. 
The SP dumbbells simulations reproduce the experimental trends qualitatively and the data semi-quantitatively even though they do 
not include long-range hydrodynamic interactions between the particles, thereby pointing to the importance of steric effects 
and collisions in setting up collective dynamics in self-propelled suspensions. The Vicsek model, while predicting qualitatively 
correct trends, leads to a rather poor quantitative agreement, perhaps due to the simplified steric description of the constituents of our system. 
Finally, our simulations show that the interplay of crowding and collective behaviour due to collisions can give rise to an intriguing non-monotonic behaviour of both the crossover time and the late time effective diffusivity of self-propelled particles. 
It would be interesting to perform similar joint theoretical and experimental analysis of suspensions of other synthetic, e.g. osmotically propelled~\cite{brady1,brady2,chucker}, microscopic swimmers.

The authors thank V. Martinez and A. N. Morozov for useful discussions and 
for a careful reading of the manuscript, and D.B.Kearns (University of Indiana) for 
providing the genetically modified strain of B. Subtilis  used in this work.
C.\ V.\ acknowledges financial support from an Intra-European Marie Curie
Fellowship. This work has made use of the resources provided by the
Edinburgh Compute and Data Facility (ECDF). The ECDF is partially
supported by the eDIKT initiative.


\begin{thebibliography}{10}

\bibitem{sriram} S. Ramaswamy, {\em Ann. Rev. Cond. Matt.}, 2010, {\bf 1}, 323.

\bibitem{menon} G. I. Menon, arXiv:1003.2032. 

\bibitem{ortiz}
J.P.Hernandez-Ortiz, C.G.Stoltz and M.D.Graham, {\em Phys.Rev.Lett.}, 2005, {\bf 95}, 204501.

\bibitem{goldstein1}
C.Dombrowski, L.Cisneros, S.Chatkaew, R.E.Goldstein and J.O.Kessler, {\em Phys.Rev.Lett}, 2004, {\bf 93}, 098103.

\bibitem{joanny} R. Voituriez, J. F. Joanny and J. Prost,
{\em Phys. Rev. Lett.}, 2006, {\bf 96}, 028102.

\bibitem{mike} M. E. Cates, D. Marenduzzo, I. Pagonabarraga and J. Tailleur,
{\em Proc. Natl. Acad. Sci. USA}, 2010, {\bf 107}, 11715. 

\bibitem{kessler1}
T.J.Pedley and J.O.Kessler, {\em Ann.Rev.Fluid Mech.}, 1992, {\bf 24}, 313.

\bibitem{Wu:2000/a}
X-L.~Wu and A.~Libchaber, {\em Phys.Rev.Lett}, 2000, {\bf 84}, 3017.

\bibitem{comment:2001/a}
G.~Gregoire, H.~Chate and Y.~Tu, {\em Phys.Rev.Lett.}, 2001, {\bf 86}, 556.

\bibitem{Vic2:1997/a}
A.~Czirok, H.~E.~Stanley and T.~Vicsek, {\em J.Phys.A}, 1997, {\bf 30}, 1375.

\bibitem{chate3:1998/a}
J.~Toner and Y.~Tu, {\em Phys.Rev.E}, 1998, {\bf 58}, 4828.

\bibitem{chate4:2001/a}
G.~Gregoire, H.~Chate and Y.~Tu, {\em Phys.Rev.E}, 2001, {\bf 64}, 011902.

\bibitem{tonertu1}
J.Toner and Y.Tu, {\em Phys.Rev.Lett}, 1995, {\bf 75}, 4326.

\bibitem{tonertu2}
J.Toner and Y.Tu, {\em Phys.Rev.E}, 1995, {\bf 58}, 4828.

\bibitem{ramaswamy1}
R.A.Simha and S.Ramaswamy, {\em Phys.Rev.Lett}, 2002, {\bf 89}, 058101.


\bibitem{ramaswamy2}
Y.Hatwalne, S.Ramaswamy, M.Rao and R.A.Simha, {\em Phys.Rev.Lett}, 2004, {\bf 92}, 118101.

\bibitem{baskaran}
A. Baskaran and M. C. Marchetti, {\it Proc. Natl. Acad. Sci. USA}, 2009, {\bf 106}, 15567.

\bibitem{isaac} I. Llopis and I. Pagonabarraga, {\em Europhys. Lett.}, 2006, {\bf 75}, 999.

\bibitem{mehandia} V. Mehandia and P. R. Nott, {\em J. Chem. Phys.}, 2008, {\bf 595}, 239.

\bibitem{BergEcoli}
H.~C.~Berg, {\it E. coli in motion}, N.Y. Springer, 2004.

\bibitem{blair}
M. A. A. Mathews, H. L. Tang and D. F. Blair, {\it J. Bacteriol.}, 1998, {\bf 180}, 5580.

\bibitem{turner}
L.Turner, W.S.Ryu and H.C.Berg,  {\it J. Bacteriol.}, 2000, {\bf 182}, 2793

\bibitem{martin}
M. Li and J. Arlt, in preparation.

\bibitem{tracking}
 J.C. Crocker and D.G. Grier, {\em J. Colloid Interface Sci.}, 1996, {\bf 179}, 298


\bibitem{Vic1:1995/a} 
T.~Vicsek, A.~Czirok, E.~Ben-Jacob, I.~Cohen and O.~Shochet, {\em Phys.Rev.Lett}, 1995, {\bf 75}, 1226.

\bibitem{dufresne}
E. R Dufresne, D Altman, D. G Grier,
{\em Europhys. Lett.}, 2001, 5{\bf 3},  264


\bibitem{LAMPPS}
S.~Plimpton, {\em J Comp Phys}, 1005, {\bf 117} 1.

\bibitem{changeloda}
C.Valeriani, A.Cacciuto, A. N. Morozov and D.Marenduzzo, in preparation.


\bibitem{gareth} 
G. P. Alexander and J. M. Yeomans, {\it Exp. Mech.}, 2010, {\bf 50}, 1283-1292.

\bibitem{baglietto}
G. Baglietto and E. V. Albano, {\em Phys. Rev. E}, 2009, {\bf 80}, 050103.


\bibitem{vincent}
L. G. Wilson, V. A. Martinez, J. Schwarz-Linek, J. Tailleur, P. N. Pusey, G. Bryant and W. C. K. Poon, {\em Phys.Rev.Lett}, 2011, {\bf 106}, 18101.

\bibitem{pedley}
T. Ishikawa, J. T. Locsei and T. J. Pedley, {\it J. Fluid Mech.}, 2008, {\bf 615}, 401.

\bibitem{pedley2}
T. Ishikawa, T. J. Pedley, {\it Phys. Rev. Lett.}, 2008, {\bf 100}, 088103.



\if{\bibitem{microbiology} M. Madigan, J. Martinko (Editors), {\it Brock Biology of Microorganisms (11th ed.)} Prentice Hall (2005).

\bibitem{bray} D. Bray, {\it Cell movements: from molecules to motility, 2nd ed.}., Garland Publishing, New York (2001).


\bibitem{chandra:1943/a}
S.~Chandrasekhar, {\em Rev.Mod.Phys.}, 1943, {\bf 15}, 1.

\bibitem{ramsha:1990/a}
R.~Ramshankar, D.~Berlin and J.~P.~Gollub, {\em Phys.Fluids A}, 1990, {\bf 2}, 1995.

\bibitem{siggia:2000/a}
B.~I.~Shraiman and E.~D.~Siggia, {\em Nature}, 2000, {\bf 405}, 639.

\bibitem{kosterlitz:2004/a}
G.~Gregoire and H.~Chate, {\em Phys.Rev.Lett.}, 2004, {\bf 92}, 25702.

\bibitem{chate2008}
H. Chate, F. Ginelli, G. Gregoire, F. Peruani and F. Raynaud, 
{\it Eur. Phys. J. B}, 2008, {\bf 64}, 451.

\bibitem{allen}
M. P. Allen and D. J. Tildeslay, {\em Computer simulations of liquids}, Oxford University Press, Oxford (1987).

\bibitem{binder:2003/a}
K.Binder, {\em Applications of Monte carlo methods to Statistical Physics}, 2003, {\bf 5}, 60.

\bibitem{brownian}
In order to recover the Gaussian distribution with zero-mean $\langle R(t) \rangle =0)$ 
 needed to satisfy a Brownian Dynamics in both $x$ and $y$ directions, we multiply by a factor of $\sqrt{3}$ in every direction of motion.}\fi

\bibitem{goldstein2010} K. Drescher, R. E.  Goldstein, N. Michel, M. Polin and I. Tuval, {\it Phys. Rev. Lett.}, 2010, {\bf 105}, 168101.

\bibitem{liverpool} R. Golestanian, T. B. Liverpool and A. Ajdari, {\it Phys.
Rev. Lett.}, 2005, {\bf 94}, 220801.

\bibitem{brady1} U. M. Cordova-Figueroa, and J. F. Brady,
{\it Phys. Rev. Lett.}, 2008, {\bf 100}, 15.

\bibitem{brady2} J. F. Brady, {\it J. Fluid Mech.}, 2011, {\bf 667}, 216.

\bibitem{chucker} C. Valeriani, R. J. Allen and D. Marenduzzo, {\it J. Chem. 
Phys.}, 2010, {\bf 132}, 204904.

\end{thebibliography}
\end{document}